# Revealing diatom-inspired materials multifunctionality

L. Musenich[a], D. Origo[b], F. Gallina[b], M. J. Buehler[c], F. Libonati[a,*]

[a] University of Genoa, Department of Mechanical, Energy, Management and Transportation Engineering, Via all'Opera Pia 15/A, 16145, Genova, Italy
[b] Polytechnic of Milan, Milano, Italy
[c] Laboratory for Atomistic and Molecular Mechanics, Massachusetts Institute of Technology, 77 Massachusetts Avenue, Cambridge, Massachusetts 02139, United States

*Corresponding author: flavia.libonati@unige.it

## ABSTRACT

Diatoms have been described as "nanometer-born lithographers" because of their ability to create sophisticated three-dimensional amorphous silica exoskeletons. The hierarchical architecture of these structures provides diatoms with mechanical protection and the ability to filter, float, and manipulate light. Therefore, they emerge as an extraordinary model of multifunctional materials from which to draw inspiration. In this paper, we use numerical simulations, analytical models, and experimental tests to unveil the structural and fluid dynamic efficiency of the *Coscinodiscus* species diatom. Then we propose a novel 3D printable multifunctional biomimetic material for applications such as porous filters, heat exchangers, drug delivery systems, lightweight structures, and robotics. Our results demonstrate Nature's role as a material designer for efficient and tunable systems and highlight the potential of diatoms for engineering materials innovation. Additionally, the results reported in this paper lay the foundation to extend the structure-property characterization of diatoms.

## INTRODUCTION

Multifunctional materials exploration has emerged as a strategy to overcome product performance limitations resulting from design tradeoffs. Consider, for instance, scenarios where catastrophic failure is not permissible, when dealing with nuclear pressure vessels, bridges, ships, medical prosthetics, or aircraft engines. Their structural parts are typically made of materials possessing moderate strength yet high fracture resistance (i.e., toughness). In these cases, the availability of multifunctional materials capable of simultaneously exhibiting these two mutually exclusive properties, such as strength and toughness, thus featuring "damage tolerance", stands as a pivotal goal with the potential of enhancing efficiency, safety, and sustainability[1,2]. Multifunctional materials conceptualization extends beyond the mere combination of diverse structural functions: it encompasses materials with integrated structural and non-structural capabilities[3]. Specifically, multifunctional materials can be engineered to manifest unusual mechanical attributes, coupled with electrical, magnetic, optical, thermal, self-healing, and other functionalities that may work synergistically or sequentially over time to yield advantages that surpass the sum of the individual ones[3,4]. While composite materials have traditionally been designed to gain functional advantages over isotropic homogeneous materials[3,5], new enabling technologies, such as additive manufacturing, have expanded design possibilities, leading to innovative examples of multifunctional meta- and 4D materials[4,6–8]. Yet, the palette of available multifunctional materials remains limited due to the complex characterization they require[9,10].

Biomimicry emerges as a captivating solution to this challenge[11–14]. Naturally occurring materials, having evolved over billions of years into complex multiscale structures optimized for simultaneously performing multiple functions, offer invaluable insights. Emulating their hierarchical architectures can streamline the development process of new multifunctional materials, providing fruitful guidance to craft a novel material-device with superior performance. Bone- and nacre-inspired materials serve as the most striking example of this biomimetic approach[15–22]. However, when it comes to multifunctionality, diatoms[23] take center stage. The intricate siliceous exoskeleton of these microscopic algae, known as the "frustule", not only acts as a robust physical barrier resistant to damage but also controls nutrient acquisition, sink rate, and in-stream diatom transport, act as a filter against viruses, and manages light absorption for the organism's self-sustainment, thus representing an outstanding model of multifunctional materials [24–26]. Diatom frustules have been studied experimentally through imaging techniques[27–29] and mechanical tests at different length scales[30–33], as well as using numerical techniques, ranging from molecular dynamics[34,35] to finite element (FE) analysis[32,33,36–38]. Several biomimetic models have been proposed for applications spanning from construction[39] to filtering devices[40,41], energy storage systems[42–44], tissue regeneration implants[45–47], optical devices[48–50], and drug delivery ones[51,52]. In a recent work by the co-authors[53], a biomimetic data-driven approach has also been used to unveil the crucial role of the gradient profile of the diatom frustule, leading to a novel diatom-inspired architected material with outstanding energy absorption capacity w.r.t. conventional honeycombs. Nevertheless, there is a lack of multidisciplinary studies that harness the potential of diatoms for novel multifunctional materials development.

In this work, we focus on *Coscinodiscus* species diatom frustules and propose an innovative bioinspired multifunctional material featuring an optimal blend of structural and fluid dynamic properties. Figure 1 illustrates the adopted framework. Starting from the diatom morphology, we first build a biomimetic material model and compare it with the natural structure. We scale up the model to the millimeter size[54] to analyze its bending and compressive response via experimental tests on 3D-printed samples. Then, we use numerical models to explore the relationship between its hierarchical structure and multifunctionality and propose analytical models to generalize the computational findings. Since frustule performance is significantly affected by the geometric features of its pores[31,32,37,41,55–59], our study covers about ten different geometric configurations to understand their impact. Through performance efficiency parameters, we identify the configuration that embodies the optimal combination of flexural stiffness, compressive strength, fluid dynamic properties, and lightweight design. The outcome of this study unveils the efficacy of diatom mimicry in shaping multifunctional materials that could find potential applications in porous filters, heat exchangers, drug delivery systems, lightweight structures, and robotics. Moreover, it lays the foundation for further research that delves deeper into the structure-property relationship of diatom frustules, exploring further aspects not addressed in this work, such as optical functions.

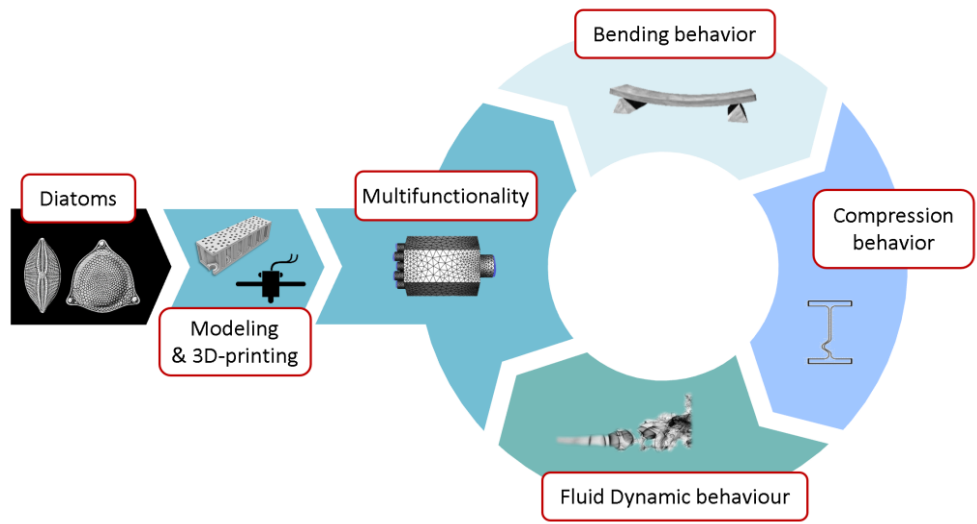

*Figure 1 - Diatom-inspired multifunctional material design. The Coscinodiscus species diatom frustule is modeled, and its structure-property relationship is characterized through experimental tests on 3D printed samples. Then numerical simulations are exploited to investigate the structural and fluid dynamic performance of the biomimetic material. The diatom picture is modified from [52], under CC BY 4.0.*

# RESULTS AND DISCUSSION

**Biomimetic model validation**

*Coscinodiscus* species diatom frustule is characterized by an intricate hierarchical structure mainly consisting of: (i) a first biomineralized plate featuring reinforced holes, known as foramen, (ii) an intermediate honeycomb-like structure called areolae, (iii) and a final biosilica porous layer named cribrum[60] (see Figure 2 (a)). Our biomimetic material model omits the irregularity associated with the morphological features of the natural structure (Figure 2 (b)). Before characterizing its multifunctionality, we validate its capability to faithfully reproduce the behavior of the mimicked biological material through FE analyses reproducing the experimental tests conducted by Aitken et al.[32].

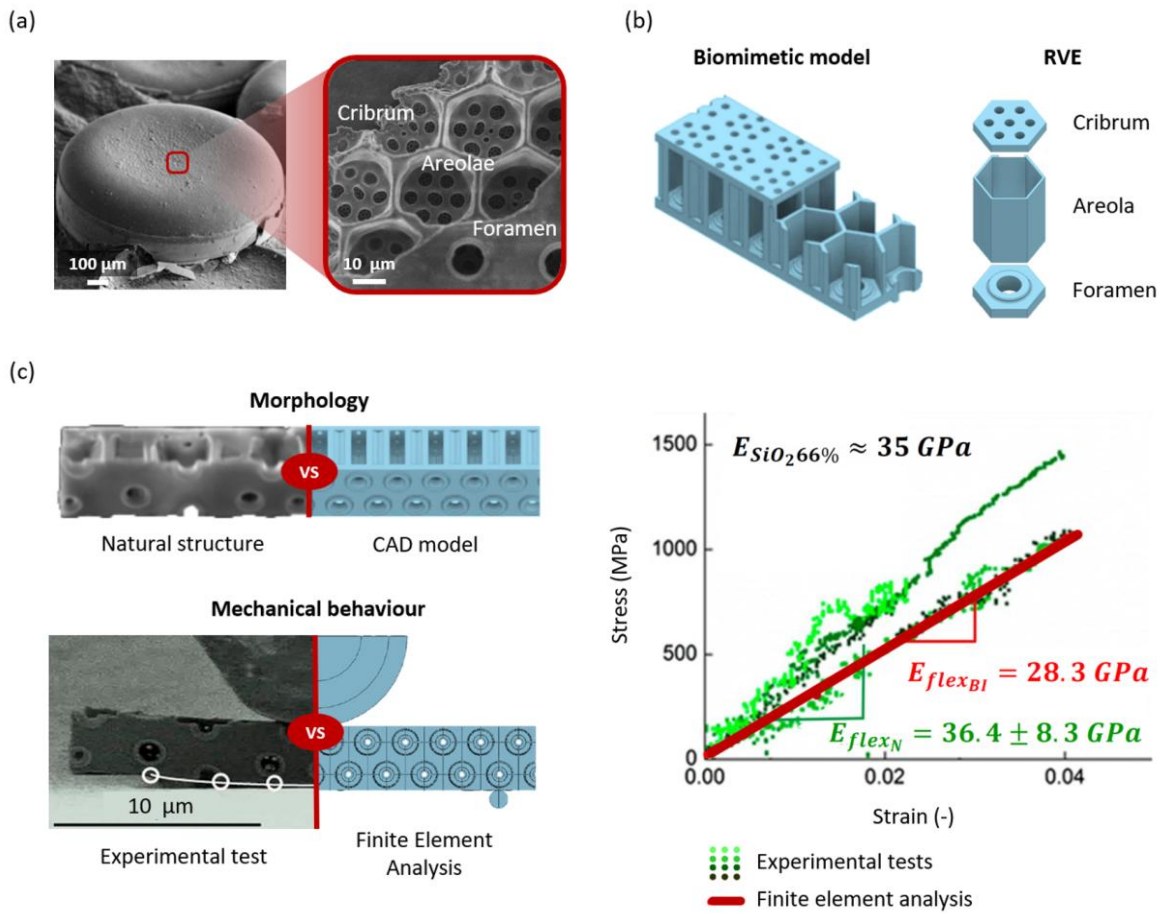

*Figure 2 - (a) Images of the diatom Coscinodiscus species according to two different magnification scales. Zooming on the halve of the frustule shows its multilayer architecture. (b) Biomimetic material model and visualization of its Representative Volume Element (RVE). (c) Comparison of natural and engineered frustule. To verify the reliability of the simplified regular geometry rebuilt by CAD, a 1:1 scale model is built, based on the study conducted by Aitken et al.[32], and its mechanical behavior under three-point bending compared, via finite element analysis, with that of the natural material analyzed by the authors.*

Figure 2 (c) shows the results obtained. The biomimetic structure exhibits a flexural elastic modulus of $E_{flex_{BI}}$ = 28.3 GPa, while the natural one has a value of $E_{flex_N}$ = 36.4±8.3 GPa. Despite the geometric differences between the two materials, the numerically calculated value falls within the range of uncertainty relative to the experimental measurement. Hence, the two structures can be considered to have similar mechanical behavior, and the biomimetic model is assumed to be representative of the natural material. On that basis, we rescale it to a dimensional scale suitable for 3D printing and study its behavior by assuming a polymeric material as the constitutive one.

**Effect of hierarchical design on frustule structural performance**

To investigate the bending performance of the bioinspired frustule, we conduct three-point bending (3PB) tests on both the multilayer biomimetic model and the single layers. Figure 3 (a) displays the load-displacement curves obtained by linearly interpolating the data from the experiments. Figure 3 (b) and Table 1 present the bending stiffness ($k_{3PB}$) and relative density ($\rho_{rel}$) of the specimens, emphasizing the impact of the lightweight design of the frustule. Relative density is defined as the ratio of their actual volume to the volume occupied by a full solid made of the same material and having the same dimensions in space.

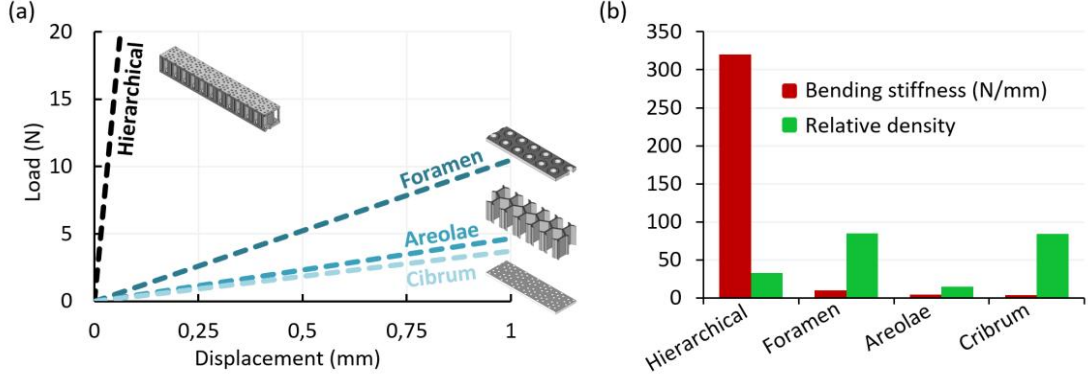

*Figure 3 - (a.) Load-displacement curves obtained from experimental three-point bending tests for the biomimetic material and its constituent substructures. (b) Bending stiffness of the tested specimens and corresponding relative density.*

*Table 1 - Results of the three-point bending experimental tests obtained for the hierarchical biomimetic model and its individual layers.*

|  | **Hierarchical** | **Foramen** | **Areolae** | **Cribrum** |
|---|---|---|---|---|
| Bending stiffness $k_{3PB}$ (N/mm) | 320 | 10.5 | 4.7 | 3.6 |
| Relative density $\rho_{rel}$ (%) | 32.9 | 85.3 | 15.2 | 84.5 |
| $\frac{k_{3PB}}{\rho_{rel}} \left(\frac{N}{mm}\right)$ | 9.73 | 0.11 | 0.31 | 0.04 |

The bioinspired hierarchical architecture allows for an enhancement in the performance of the individual design elements. Specifically, $\frac{k_{3PB}}{\rho_{rel}}$, the ratio of the flexural stiffness to relative density, is nearly two orders of magnitude higher than that of its constituent substructures. The predominant contribution to this mechanical response comes from the honeycomb-like substructure (*i.e.*, the areolae). Indeed, despite the load-displacement curve of the foramen being the stiffest among those of the individual layers, in terms of lightweight design (i.e., the $k_{bend}/\rho_{rel}$ ratio), the areolae prove to be more efficient, showing a three-fold increase in terms of $\frac{k_{3PB}}{\rho_{rel}}$. Thus, to further explore the structure-property relationship of the biomimetic frustule, we focus on the geometrical features of this layer.

*Role of the honeycomb-like layer*

Figure 4 (a) shows the bioinspired structure bending stiffness estimated through FE analysis compared to its analytically calculated flexural moment of inertia, as the thickness of the areolae walls varies. The moment of inertia consistently increases as the areolae wall thickness decreases due to the material being relocated from the hexagonal cells to the two outer layers. When assimilating the frustule section to an I-shaped profile beam, this trend aligns with expectations. What is surprising compared to that is the trend of the bending stiffness. Indeed, while the beam theory suggests a direct proportionality between the two quantities involved, the results reveal a different relationship. Specifically, by analyzing the course of bending stiffness, two distinct regions can be distinguished. In the region marked by a positive increase in areolae wall thickness (shaded green), the bending stiffness appears to be minimally affected by the redistribution of material in the different layers of the biomimetic model. This response is primarily associated with the heterogeneity of the frustule section, not considered in the analytical function used (described in the Supporting Information (SI) file), which assumes a homogeneous cross-section, according to Euler-Bernoulli hypothesis. On the other hand, when the areolae wall thickness becomes smaller, the bending stiffness decreases as the moment of inertia rises. The most plausible hypothesis justifying this behavior is that buckling occurs in the areolae, leading to a zeroing of the out-of-plane stiffness of the hexagonal cells.

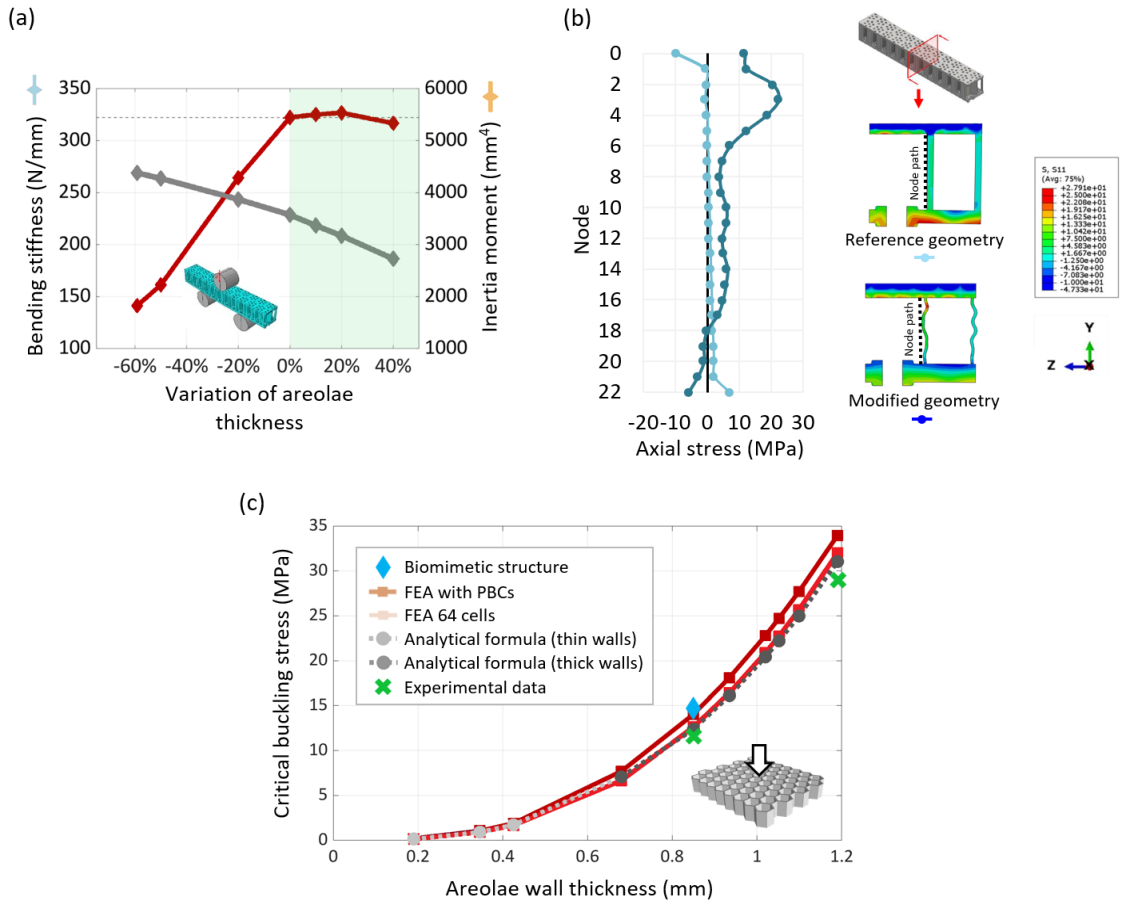

*Figure 4 - (a) Biomimetic frustule bending stiffness and inertia moment variations according to the areolae wall thickness values. (b) Axial stress distribution in the most loaded section of the frustule when subjected to three-point bending, according to the original geometry and to a modified configuration with thinner areolae walls. (c) Trend of critical buckling stress as the wall thickness of the areolae changes, evaluated for out-of-plane compression. The graph also shows the value calculated for the entire biomimetic frustule.*

To validate the buckling hypothesis, we initially assess the stresses occurring in the most loaded section of the biomimetic model when subjected to three-point bending. Figure 4 (b) presents the distribution of stresses along its principal axis (i.e., the x-axis depicted in the figure), both for the model created initially by rescaling the size of the natural frustule and for a varied configuration in which the wall thickness of the areolae is reduced by 60%. The modified one exhibits a more irregular stress distribution compared to the natural counterpart, supporting the hypothesis of failure due to elastic instability. Furthermore, this evidence underscores how Nature has optimized the distribution of the material in the *Coscinodiscus* species diatom frustule to maximize its mechanical performance. These findings lead us to delve into the buckling phenomenon by examining the out-of-plane compression behavior of the areolae. Figure 4 (c) illustrates the variation of its critical stress as a function of the honeycomb wall thickness. Numerical, analytical, and experimental data exhibit excellent agreement, confirming the cubic dependence of the buckling critical load on the thickness of the cellular walls, as predicted by the theory[61]. Furthermore, these results elucidate the observed anomaly in the flexural stiffness trend reported in Figure 4 (a). In the same plot, we also present the buckling stress value for the entire biomimetic frustule in its originally derived configuration. Compared to the individual layer, the hierarchical structure demonstrates a higher strength. Nevertheless, the difference from the experimental data remains below 15%, confirming the primary role of the areolae in the mechanical behavior of the diatom. The analytical model, while conservative, can be used with the understanding that it is conservative with respect to the behavior of the whole biomimetic structure. In contrast, the FE model, composed of a single hexagonal cell with periodic boundary conditions, seems to correctly represent the behavior of the frustule, suggesting that it can be effectively used to analyze the frustule compressive behavior as its geometric parameters change. Additionally, it is worth noting that the experimental results are influenced by inherent anisotropies and imperfections introduced by the additive manufacturing process. These aspects, omitted in our simulations, justify the observed differences in results.

*Role of porous layers*

To evaluate the mechanical role of the cribrum and foramen, we analyze the structure-properties relationship of these two frustule porous substructures following three distinct strategies, as depicted in Figure 5 (a). To mitigate 3PB localized effects, we perform four-point bending (4PB) analysis. Figure 5 (b) shows the results obtained in terms of absolute bending stiffness ($k_{4PB}$) under 4PB loading and $\eta_{flex}$, the bending stiffness normalized to the model mass (m):

$$\eta_{flex} = \frac{k_{4PB}}{m} \qquad (1)$$

A higher value of $\eta_{flex}$ indicates better bending performance per unit weight, serving as a measure of flexural efficiency. The foramen, being thicker than the cribrum, is the stiffer layer. Its most efficient geometric configuration appears to be without the rings and with a volume equal to the originally developed biomimetic model. However, the latter is not far behind, exhibiting an $\eta_{flex}$ equal to 94% of the best case. Since the presence of rings around the holes benefits crack propagation arrest and material damage-tolerance[32], the naturally rescaled foramen emerges as the optimal structure. In contrast, no significant differences in terms of performance or structural functions are observed among the cribrum cases. Once again, the best geometric model is the regular one. Although variations in $\eta_{flex}$ across configurations are modest (~5-10%), their irregular geometry poses disadvantages in terms of 3D modeling of the biomimetic material and manufacturing processes.

**Effect of hierarchical design on frustule fluid dynamic performance**

The areolae, with its honeycomb architecture, represents the most efficient geometric configuration that Nature can make to achieve maximum area with minimum perimeter[62]. In the case of the *Coscinodiscus* diatom species, this translates into the capacity of storing more nutrients into a larger volume while minimizing the energy consumption required to build a structure to contain them. By contrast, the role of the frustule porous layers in facilitating the assimilation of nutrients and the expulsion of waste remains a subject of inquiry. To gain insights into this aspect, we conduct computational fluid dynamics (CFD) analyses. Specifically, we simulate the behavior of the frustule representative volume element (RVE) as its geometric configuration changes (see Figure 5 (a)), initially assuming a flow from the cribrum to the foramen (inward), and subsequently, from the foramen to the cribrum (outward). Models named "freePosVx" represent three geometric configurations with randomly varied hole positions, while variants with free cribrum hole diameters are labeled "freeDimVx". The models labeled "NOR" and "NOR_sameVol" lack reinforcement around the foramen hole. The former is obtained by simply removing the reinforcement ring from the original geometric model (no ring – NOR). In the second, the volume of material corresponding to the reinforcement ring is redistributed into the foramen thickness (NOR_sameVol). Table S1 provides detailed results on the flow characteristics analyzed numerically. Considering the inward case, only the distribution of the cribrum holes seems to significantly influence the average behavior of the flow, while their size locally alters the velocity profile without globally affecting the overall flow. These variations can be justified by the continuity equation for incompressible flows, and no remarkable or unexpected phenomena emerge. However, from the visualization of the flow lines (Figure 5 (c)), the flow redistribution effect associated with the presence of reinforcing rings around the foramen holes becomes apparent. In the absence of rings, the portion of the flow that fails to enter the foramen holes hits its surface, and the flow particles collect near the hole edges. Conversely, the presence of rings creates retrograde flows, allowing the material that initially failed to enter the foramen hole to later enter as a second "pulse". This observation reinforces the hypothesis that *Coscinodiscus* species diatom is particularly efficient in aquatic environments where nutrients are not evenly distributed, necessitating storage in areas of higher concentration to survive in regions with lower nutrient availability[56]. Analyzing the outward flow, on the other hand, the primary effect associated with the variation in geometric features of the biomimetic frustule lies in the tortuosity of the flow[63], leading to a global reduction in velocity. Additionally, the regularity of the original RVE promotes a more uniform distribution of pressure within the frustule "chambers", preventing local peaks that could induce critical stresses on the biological material (see Figure 5 (d)).

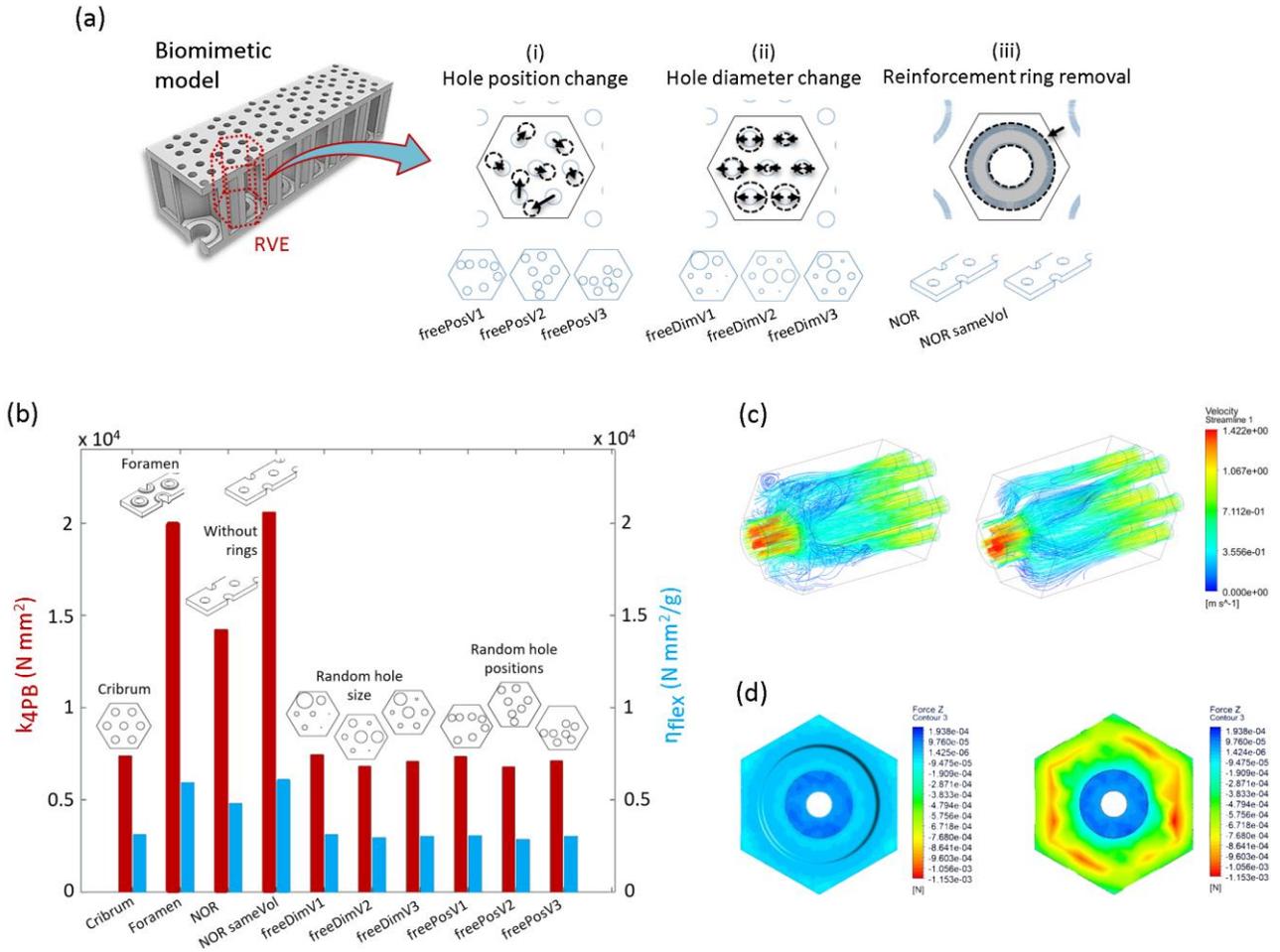

*Figure 5 - (a) Exploration of geometric variations in biomimetic porous layers to enhance material functionality. Strategies include (i) randomly relocating holes in the cribrum, (ii) randomly varying the hole sizes, and (iii) assessing the effect of the foramen reinforcement ring. Visualization of flow patterns and axial force distribution with and without reinforcement rings. (b) Bending stiffness ($k_{4PB}$) under four-point bending loading and the corresponding flexural efficiency ($\eta_{flex}$) across varying geometric configurations of the biomimetic frustule. (c-d) Visualization of flow patterns (c) and axial force distribution (d) with (left) and without (right) foramen reinforcement rings.*

**Biomimetic material multifunctionality**

Figure 6 shows radar plots to assess the biomimetic frustule multifunctionality. Since lightweight design is one of the main goals and advantages of the diatom-inspired architecture, here we use the previously defined flexural efficiency $\eta_{flex}$ as a benchmark for evaluating the frustule structural behavior. Additionally, since buckling is a predominant failure mode in diatom mechanics, we also report the related critical stress $\sigma_{buckl}$. For fluid dynamic performance, we introduce a fluid dynamic efficiency parameter, $\eta_{fluid}$, representing the filtering or waste removal capacity of the frustule, and a flow distribution efficiency parameter, $\eta_{force}$, accounting for the failure risk associated with flow pressure. These two parameters, specifically, are defined as follows:

$$\eta_{fluid} = \frac{\dot{m}_{outlet,out}}{\dot{m}_{outlet,in}} \quad (2)$$

where $\dot{m}_{outlet,out}$ and $\dot{m}_{outlet,in}$ are the outgoing mass flow rate under the outward and inward flow conditions, respectively, and

$$\eta_{force} = \frac{1}{F_{max_{foramen}}} \quad (3)$$

where $F_{max_{foramen}}$ is the maximum force produced by the inward fluid on foramen. In all reported cases, the critical buckling stress is maximum as it mainly depends on the geometric features of the areolae. For the cribrum, the original pore proportions and design ensure the best multifunctional performance. The "freeDimV2" case stands as an exception, proving to be more efficient in distributing the acting pressure on the diatom frustule. However, given that the difference with the baseline case is minimal, it is considered an outlier here. The exploration of the reasons behind this evidence is left to future studies. The geometric characteristics of the reference model are also advantageous for the foramen. The presence of reinforcing rings significantly enhances the multifunctional efficiency of the diatom-inspired material model.

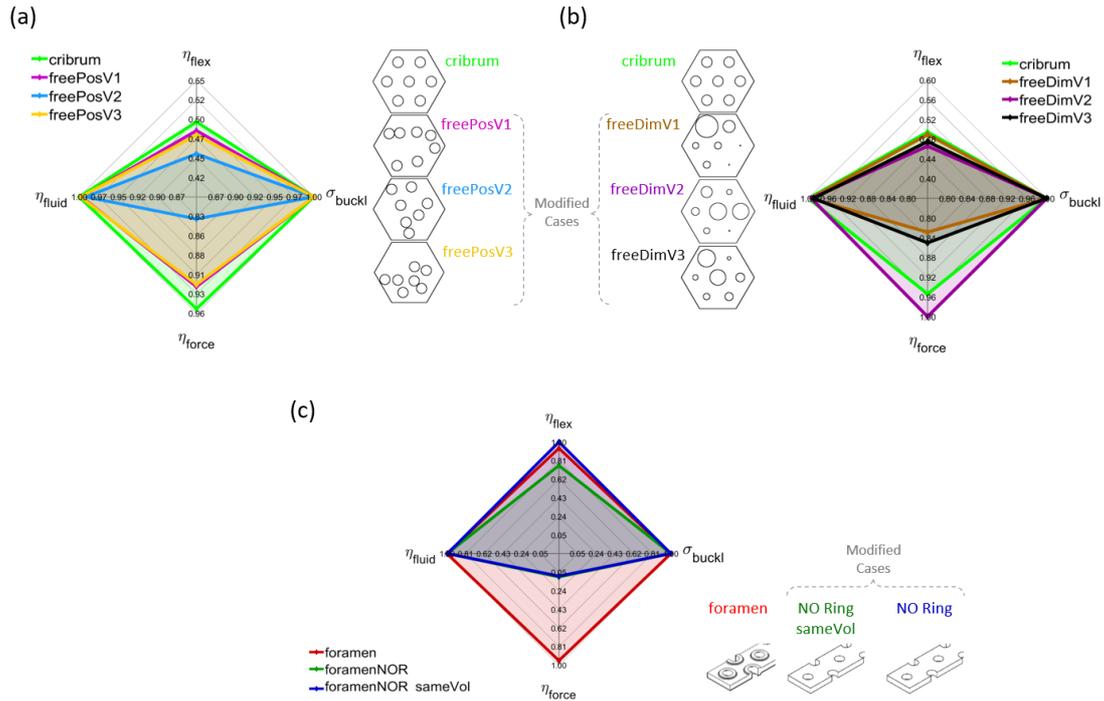

*Figure 6 - Radar plot assessing multifunctionality in the diatom frustule-inspired materials, measuring mechanical ($\sigma_{buckl}$, $\eta_{force}$, $\eta_{flex}$) and fluid dynamic ($\eta_{fluid}$) performance. (a) Effect of varied hole positions. (b) Effect of varied hole sizes. (c) Effect of foramen reinforcement ring.*

To simplify the quantification of the differences between the various designs analyzed in terms of multifunctionality, we report in Table 2 and Table 3 the values of multifunctional efficiency $\eta_{multi}$:

$$\eta_{multi} = \eta_{flex} + \eta_{fluid} + \eta_{force} \quad (4)$$

and its percentage differences ($\Delta\eta_{multi\%}$) compared to the reference geometry, respectively for the cribrum layer and the foramen. As observed, the difference in $\eta_{multi}$ between the "freeDimV2" case and the original model is less than 3%. For the foramen, instead, the benefits derived from the presence of the reinforcing ring are quite evident, and even in the worst case, the original geometry is significantly more efficient (+26%).

*Table 2 - Multifunctional efficiency of cribrum geometric models and differences between the varied geometries and the original one.*

| Geometric model | $\eta_{multi}$ | $\Delta\eta_{multi\%}$ |
|---|---|---|
| Cribrum | 2.4 | |
| freePosV1 | 2.4 | -0.6 |
| freePosV2 | 2.3 | +4.8 |
| freePosV3 | 2.4 | -0.2 |
| freeDimV1 | 2.3 | +3.2 |
| freeDimV2 | 2.5 | -2.9 |
| freeDimV3 | 2.3 | +3.1 |

*Table 3 - Multifunctional efficiency of foramen geometric models and differences between the varied geometries and the original one.*

| Geometric model | $\eta_{multi}$ | $\Delta\eta_{multi\%}$ |
|---|---|---|
| Foramen | 2.8 | |
| NOR | 1.9 | +35 |
| NOR sameVol | 2.1 | +26 |

## CONCLUSIONS

In this study, we present a novel model of multifunctional materials that faithfully reproduces the geometry and functionality of the exoskeleton of the *Coscinodiscus* species diatom. We explore its structure-property relationship through numerical simulations, analytical models, and experimental tests conducted on 3D-printed samples. Our investigation highlights both the structural efficiency of its hierarchical design in resisting bending and compression loads and its fluid dynamic efficiency in interacting with multidirectional flows, elucidating the role of its individual constituent layers in determining these performances. The regular structure, obtained by idealizing the biological frustule, proves to be advantageous, amplifying the properties of its design elements and outperforming the varied geometric configurations analyzed. It represents a high-potential model for multifunctional materials in applications requiring an optimal combination of structural and fluid dynamic properties, such as porous filters, heat exchangers, drug delivery systems, or robotic systems. For instance, by exploiting the ability to induce elastic instabilities in the frustule by controlling the thickness of the areolae walls, structural collapse could be harnessed to design new flow control valves, where material and structure are integrated. Alternatively, by controlling the porosity of various layers, it could be employed in catalytic systems or scaffolds for tissue regeneration. The biomimetic approach we propose here involves the use of polymeric materials to fabricate frustule samples, altering the properties associated with its bio-ceramic nature. Thus, to extend the multifunctionality of the proposed material, it becomes necessary to investigate the effect of the underlying constituent material of this hierarchical architecture. A promising approach to expedite this process is the application of Materials Informatics techniques. The diversity of diatom shapes and properties observed in Nature could facilitate the construction of datasets to build predictive models[64] and further optimize the bioinspired material multifunctionality.

# MATERIALS AND METHODS

**Material design approach**

A modular approach for biomimetic material modeling is adopted, starting with the definition of an RVE, which is then replicated in a 2D space to create a periodic cellular material. The design of the frustule RVE is based on geometric data from natural diatoms, detailed in the SI file. Initially, a 1:1 scale model is generated to assess its fidelity in reproducing the biological structure's behavior. Subsequently, it is scaled up to the millimeter size to be suitable the additive manufacturing process for sample production. To explore the structure-property relationship (see Figure 5 (a)), MATLAB scripts are created to develop geometries with variables pore position and size. Results are normalized to mass since volume constancy is not specified.

**3D printing**

Samples are produced employing an Objet500 Connex1, a multi-material polyethylene-jet 3D printer by Stratasys. VeroBlack® is used for the biomimetic structure's architecture, while a water-soluble wax-like material generates supports for cantilever parts, subsequently removed with a water jet (refer to Figure 7 (a)). Both multilayer and monolayer specimens are 3D printed (three for each configuration) at a 5000x magnification compared to natural diatoms (see Figure 7 (b)). Additionally, to establish the material model for numerical simulations, dog bone samples are 3D printed in VeroBlack® following the ASTM D638-14 standard.

**Mechanical testing**

3PB tests are conducted using an MTS Synergie testing machine with a 1 kN load cell, implementing displacement control at a velocity of 2 mm/min. The experimental setup involves placing full-geometry 3D-printed samples on the machine to replicate a predatory attack on the diatom frustule (see Figure 7 (c)). Support pins, represented by two steel cylinders with a diameter of 20 mm (like the punch), are spaced 60 mm apart. For 4PB experiments, the ASTM D6272-17 standard is followed, utilizing displacement control mode at a speed of 1 mm/min. Compression tests on the honeycomb-like layer are performed according to the ASTM C365/C365M-16 standard, employing an MTS Alliance RF150 machine with a 150 kN load cell (Figure 7 (c)). Tests are conducted in displacement control mode at a crosshead speed of 0.5 mm/min on honeycombs with two different wall thickness values to evaluate their impact on the material's out-of-plane mechanical behavior. Lastly, tensile tests are performed on dog-bone samples to determine the elastic properties of VeroBlack®. These experiments are conducted according to ASTM standard D638-14, utilizing an MTS Alliance RT100 universal testing machine equipped with a 100 kN load cell.

**Numerical simulations**

*Structural analyses*

The validation of the biomimetic material model is conducted by simulating the experimental conditions of Aitken et al.'s tests[32] using Abaqus CAE 2017. Linear elastic models are employed to represent the brittle behavior of biosilica and bending test fixtures. For the beam, Young's modulus is set to 35 GPa with a Poisson's coefficient of 0.17, for the punch, Young's modulus is 1000 GPa with a Poisson's coefficient of 0.2, and for the supports, Young's modulus is 200 GPa with a Poisson's coefficient of 0.27. A structured mesh of hexahedral elements is utilized due to geometric regularity, resulting in approximately 850,000 finite elements. A sensitivity analysis of the mesh is also performed to ensure the results are not mesh-dependent. The simulation, under displacement control, aims to mimic the experimental deformation with a 1 nm vertical displacement assigned to the punch, calculating the reaction force of the beam directly on the punch. Standard surface-to-surface contact with frictionless behavior is selected between the punch and the beam, and generic contact with a friction coefficient of 0.1 is chosen between the beam and the steel supports to represent friction. To replicate the mechanical tests on 3D-printed biomimetic material samples, a linear elastic model of VeroBlack is used, characterized by a Young's modulus of 1567 MPa and a Poisson's coefficient of 0.35 (obtained from tension tests, see SI). Mesh and boundary conditions are adapted based on reference mechanical tests. Rigid bodies are used to model the parts of the testing machine in contact with the specimens for 3PB tests, utilizing surface-to-surface contacts with a friction coefficient of 0.15 (Figure 7 (d)). The same approach is applied to 4PB tests. For buckling analysis, equivalent static actions replace the punch and pins. Hexagonal-dominated elements with an overall size of 0.5 mm are used for meshing. For compression analyses, a single honeycomb cell is modeled to optimize computational cost, with constraints on displacements and periodic boundary conditions[65]. Quad-dominated shell elements with an average size of 0.4 mm are used for meshing. Model accuracy is assessed by comparing it with a honeycomb model characterized by 64 cells.

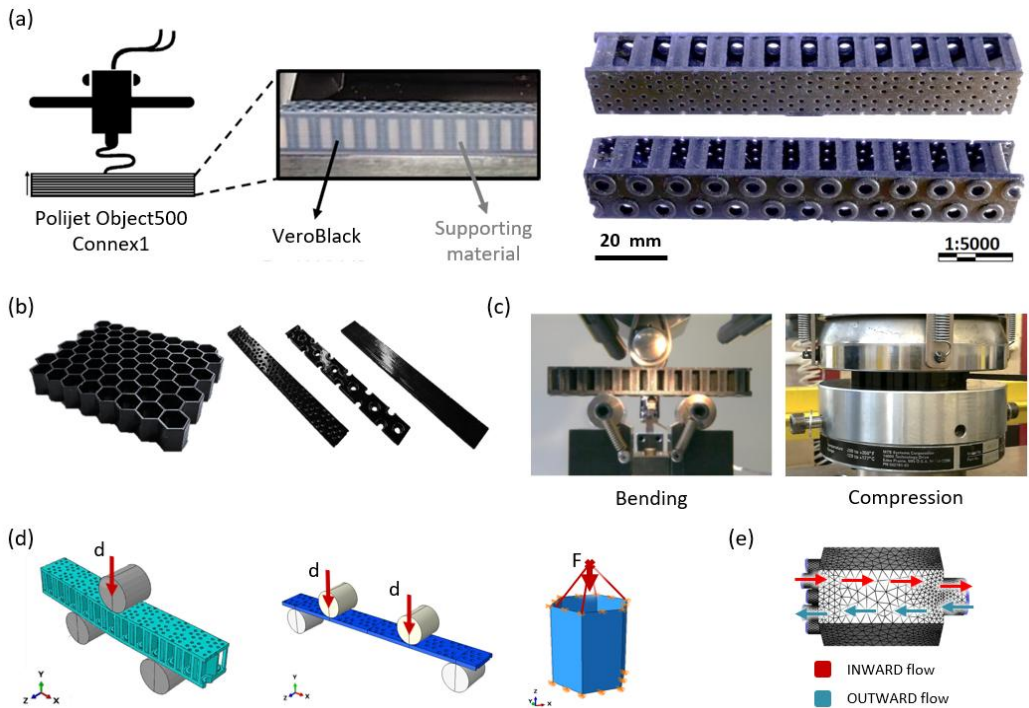

*Figure 7 - (a) Additive manufacturing process used to make samples of the diatom-inspired material. (b) 3D printed samples of the material component layers. (c) Mechanical tests performed to understand the biomimetic frustule structural behavior. (d-e) Numerical simulations conducted in this study: (d) Finite element analyses to assess the structural behavior–under flexural and compressive loading– of the biomimetic material and its substructures; (e) Computational Fluid Dynamics (CFD) simulations to assess the fluid dynamic efficiency regarding the filtering and redistribution capacity of the flow within the biomimetic material.*

*Computation fluid dynamic analyses*

To study the flow characteristics interacting with the diatom frustule, its RVE is recreated "in negative" using CAD tools. Subsequently, the ANSYS 2019 R2 software package is employed for geometry discretization (approximately 300,000 elements), analysis, and post-processing of results (Figure 7 (e)). In the RVE inlet section, we impose a flow velocity (1 m/s) normal to the surface of the valve and parallel to the axis of the hexagonal cell. Atmospheric pressure is set in the other section, and areas not belonging to the outlet or inlet are designated as stationary static walls with a no-slip shear condition. Water is selected as the material model, utilizing properties from the ANSYS Fluent database: a density of 998.2 kg/m³ and viscosity of 0.001003 kg/(m s). The flow analysis includes calculations for global average and maximum velocities, average velocity, mass flow rate at the outlet, and visualization of flow lines. The results are presented as relative/normalized values, not in terms of absolute values, as they do not represent specific experimental or natural situations.

**Analytical modeling**

*Frustule bending moment of inertia*

Due to the intricate geometry of the biomimetic model, an analytical evaluation of its bending stiffness is conducted by establishing an approximate function to calculate the bending moment of inertia. Comprehensive details regarding the derivation of this formula can be found in the SI file.

*Areolae out-of-plane compression*

The out-of-plane compressive behavior of honeycombs has been extensively studied in the literature[61,66,67]. Formulas for their critical buckling stress are derived on the premise that individual cells can be treated as a combination of plate elements, connected through vertical edges, and strongly depend on the ratio of their thickness-to-height ($t/h$). For thin-walled honeycombs ($t/h>0.1$), they are often treated as thin plates. However, in cases where the wall thickness is significant or unit cells are small, shear effects become non-negligible, requiring the application of thick plate theories. To provide a unified formula for such scenarios, this paper employs the equation proposed by Piscopo[68]:

$$\sigma_{cr} = Kf \frac{E_{bulk}}{1-v^2} \left(\frac{t}{l}\right)^3 \qquad (4)$$

where $E_{bulk}$ is the Young's modulus of the material constituting the honeycomb, $t$ is the thickness of its walls, $l$ the side of the hexagon defining the bases of its elementary cells, $f$ is a correction factor to allow for shear effects, $K$ is a coefficient dependent on the loading configuration and geometric characteristics of the honeycomb, and $v$ the Poisson's ratio of its constituent material. To best fit the analytical results to the other data, we use the proposed formula with a $K$ value of 2.3. This choice reflects a condition where unloaded walls of hexagonal cells have no notable constraints, making the selected value empirical and applied retrospectively.

## ACKNOWLEDGMENTS

The authors acknowledge the support from the MISTI seed grant. F.L. acknowledges the support from the University of Genoa under the Curiosity-driven starting grant. M.J.B. acknowledges support from the Army Research Office (W911NF2220213).

## REFERENCES

[1]     Ritchie R O 2011 The conflicts between strength and toughness *Nat. Mater.* **10** 817–22
[2]     Launey M E and Ritchie R O 2009 On the fracture toughness of advanced materials *Adv. Mater.* **21** 2103–10
[3]     Ferreira A D B L, Nóvoa P R O and Marques A T 2016 Multifunctional Material Systems: A state-of-the-art review *Compos. Struct.* **151** 3–35
[4]     Kuang X, Roach D J, Wu J, Hamel C M, Ding Z, Wang T, Dunn M L and Qi H J 2019 Advances in 4D Printing: Materials and Applications *Adv. Funct. Mater.* **29** 1–23
[5]     Ali A and Andriyana A 2020 Properties of multifunctional composite materials based on nanomaterials: a review *RSC Adv.* **10** 16390–403
[6]     Yuan X, Chen M, Yao Y, Guo X, Huang Y, Peng Z, Xu B, Lv B, Tao R, Duan S, Liao H, Yao K, Li Y, Lei H, Chen X, Hong G and Fang D 2021 Recent progress in the design and fabrication of multifunctional structures based on metamaterials *Curr. Opin. Solid State Mater. Sci.* **25** 100883
[7]     Al-Ketan O and Abu Al-Rub R K 2019 Multifunctional Mechanical Metamaterials Based on Triply Periodic Minimal Surface Lattices *Adv. Eng. Mater.* **21** 1–39
[8]     Wu W, Hu W, Qian G, Liao H, Xu X and Berto F 2019 Mechanical design and multifunctional applications of chiral mechanical metamaterials: A review *Mater. Des.* **180** 107950
[9]     Olson G B 1997 Computational Design of Hierarchically Structured Materials *Science (80-. ).* **277** 1237–42
[10]    Olson G B 2000 Designing a New Material World *Science (80-. ).* **288** 993–8
[11]    Fratzl P 2007 Biomimetic materials research: What can we really learn from nature's structural materials? *J. R. Soc. Interface* **4** 637–42
[12]    Vincent J F V, Bogatyreva O A, Bogatyrev N R, Bowyer A and Pahl A K 2006 Biomimetics: Its practice and theory *J. R. Soc. Interface* **3** 471–82
[13]    Vincent J F V 2009 Biomimetics - A review *Proc. Inst. Mech. Eng. Part H J. Eng. Med.* **223** 919–39
[14]    Liu Z, Zhang Z and Ritchie R O 2018 On the Materials Science of Nature's Arms Race *Adv. Mater.* **30**

[15] Zimmermann E A and Ritchie R O 2015 Bone as a Structural Material *Adv. Healthc. Mater.* **4** 1287–304
[16] Libonati F, Vellwock A E, Ielmini F, Abliz D, Ziegmann G and Vergani L 2019 Bone-inspired enhanced fracture toughness of de novo fiber reinforced composites *Sci. Rep.* **9** 1–12
[17] Libonati F, Gu G X, Qin Z, Vergani L and Buehler M J 2016 Bone-Inspired Materials by Design: Toughness Amplification Observed Using 3D Printing and Testing *Adv. Eng. Mater.* **18** 1354–63
[18] Tan G, Zhang J, Zheng L, Jiao D, Liu Z, Zhang Z and Ritchie R O 2019 Nature-Inspired Nacre-Like Composites Combining Human Tooth-Matching Elasticity and Hardness with Exceptional Damage Tolerance *Adv. Mater.* **31** 1–9
[19] Espinosa H D, Rim J E, Barthelat F and Buehler M J 2009 Merger of structure and material in nacre and bone - Perspectives on de novo biomimetic materials *Prog. Mater. Sci.* **54** 1059–100
[20] Gu G X, Libonati F, Wettermark S D and Buehler M J 2017 Printing nature: Unraveling the role of nacre's mineral bridges *J. Mech. Behav. Biomed. Mater.* **76** 135–44
[21] Pro J W and Barthelat F 2019 The fracture mechanics of biological and bioinspired materials *MRS Bull.* **44** 46–52
[22] Wegst U G K, Bai H, Saiz E, Tomsia A P and Ritchie R O 2015 Bioinspired structural materials *Nat. Mater.* / **14**
[23] Theriot E, Herbarium D, Round F E, Crawford R M and Mann D G 1992 The Diatoms. Biology and Morphology of the Genera. *Syst. Biol.* **41** 125
[24] De Tommasi E, Gielis J and Rogato A 2017 Diatom Frustule Morphogenesis and Function: a Multidisciplinary Survey *Mar. Genomics* **35** 1–18
[25] Mishra M, Arukha A P, Bashir T, Yadav D and Prasad G B K S 2017 All new faces of diatoms: Potential source of nanomaterials and beyond *Front. Microbiol.* **8** 1–8
[26] Zhang D Y, Wang Y, Cai J, Pan J F, Jiang X G and Jiang Y G 2012 Bio-manufacturing technology based on diatom micro- and nanostructure *Chinese Sci. Bull.* **57** 3836–49
[27] Zgłobicka I, Gluch J, Liao Z, Werner S, Guttmann P, Li Q, Bazarnik P, Plocinski T, Witkowski A and Kurzydlowski K J 2021 Insight into diatom frustule structures using various imaging techniques *Sci. Rep.* **11** 1–10
[28] Darouich O, Baaziz W, Ihiawakrim D, Hirlimann C, Spehner D, Schultz P, Poncet H, Rouchon V, Labidi S, Petit C, Levitz P and Ersen O 2022 3D multiscale analysis of the hierarchical porosity in Coscinodiscus sp. diatoms using a combination of tomographic techniques *Nanoscale Adv.* **4** 1587–98
[29] Manoylov K and Ghobara M 2021 Introduction for a Tutorial on Diatom Morphology *Diatom Morphogenesis* (Wiley) pp 1–18
[30] Xu H, Shi Z, Zhang X, Pang M, Pan K and Liu H 2021 Diatom frustules with different silica contents affect copepod grazing due to differences in the nanoscale mechanical properties *Limnol. Oceanogr.* **66** 3408–20
[31] Hamm C E, Merkel R, Springer O, Jurkojc P, Maiert C, Prechtelt K and Smetacek V 2003 Architecture and material properties of diatom shells provide effective mechanical protection *Nature* **421** 841–3
[32] Aitken Z H, Luo S, Reynolds S N, Thaulow C and Greer J R 2016 Microstructure provides insights into evolutionary design and resilience of Coscinodiscus sp. frustule *Proc. Natl. Acad. Sci. U. S. A.* **113** 2017–22
[33] Moreno M D, Ma K, Schoenung J and Dávila L P 2015 An integrated approach for probing the structure and mechanical properties of diatoms: Toward engineered nanotemplates *Acta Biomater.* **25** 313–24
[34] Garcia A P, Sen D and Buehler M J 2011 Hierarchical silica nanostructures inspired by diatom algae yield superior deformability, toughness, and strength *Metall. Mater. Trans. A Phys. Metall. Mater. Sci.* **42** 3889–97
[35] Dimas L S and Buehler M J 2012 Influence of geometry on mechanical properties of bio-inspired silica-based hierarchical materials *Bioinspiration and Biomimetics* **7**
[36] Topal E, Rajendran H, Zgłobicka I, Gluch J, Liao Z, Clausner A, Kurzydłowski K J and Zschech E 2020 Numerical and experimental study of the mechanical response of diatom frustules *Nanomaterials* **10** 1–14
[37] Lu J, Sun C and Wang Q J 2015 Mechanical Simulation of a Diatom Frustule Structure *J. Bionic Eng.* **12** 98–108
[38] Abdusatorov B, Salimon A I, Bedoshvili Y D, Likhoshway Y V. and Korsunsky A M 2020 FEM exploration of the potential of silica diatom frustules for vibrational MEMS applications *Sensors*

*Actuators, A Phys.* **315** 112270

[39] Hamm C 2015 *Evolution of Lightweight Structures: Analyses and Technical Applications* vol 6, ed C Hamm (Dordrecht: Springer Netherlands)

[40] Hale M S and Mitchell J G 2001 Functional morphology of diatom frustule microstructures: Hydrodynamic control of brownian particle diffusion and advection *Aquat. Microb. Ecol.* **24** 287–95

[41] Losic D, Rosengarten G, Mitchell J G and Voelcker N H 2006 Pore architecture of diatom frustules: Potential nanostructured membranes for molecular and particle separations *J. Nanosci. Nanotechnol.* **6** 982–9

[42] Huang J, Wu B, Lyu S, Li T, Han H, Li D, Wang J, Zhang J, Lu X and Sun D 2021 Solar Energy Materials and Solar Cells Improving the thermal energy storage capability of diatom-based biomass / polyethylene glycol composites phase change materials by artificial culture methods *Sol. Energy Mater. Sol. Cells* **219** 110797

[43] Wu B, Lyu S, Han H, Li T, Sun H, Wang J, Li D, Lei F, Huang J and Sun D 2021 Biomass-based shape-stabilized phase change materials from artificially cultured ship-shaped diatom frustules with high enthalpy for thermal energy storage *Compos. Part B* **205** 108500

[44] Sun H, Li T, Lei F, Lyu S, Yang Y, Li B, Han H, Wu B, Huang J, Zhang C, Li D and Sun D 2021 Fast Self-Healing Superhydrophobic Thermal Energy Storage Coatings Fabricated by Bio-Based Beeswax and Artificially Cultivated Diatom Frustules *ACS Appl. Mater. Interfaces* **13** 48088–100

[45] Green D W, Goto T K, Kim K S and Jung H S 2014 Calcifying tissue regeneration via biomimetic materials chemistry *J. R. Soc. Interface* **11**

[46] Abdelhamid M A A and Pack S P 2021 Biomimetic and bioinspired silicifications: Recent advances for biomaterial design and applications *Acta Biomater.* **120** 38–56

[47] Jo Y K, Choi B H, Kim C S and Cha H J 2017 Diatom-Inspired Silica Nanostructure Coatings with Controllable Microroughness Using an Engineered Mussel Protein Glue to Accelerate Bone Growth on Titanium-Based Implants *Adv. Mater.* **29**

[48] Zaman S, Hassan M M, Hasanuzzaman M and Baten M Z 2020 Coscinodiscus diatom inspired bi-layered photonic structures with near-perfect absorptance accompanied by tunable absorption characteristics *Opt. Express* **28** 25007

[49] Xie P, Chen Z, Xu J, Xie D, Wang X, Cui S, Zhou H, Zhang D and Fan T 2019 Artificial ceramic diatoms with multiscale photonic architectures via nanoimprint lithography for CO2 photoreduction *J. Am. Ceram. Soc.* **102** 4678–87

[50] Li A, Zhao X, Duan G, Anderson S and Zhang X 2019 Diatom Frustule-Inspired Metamaterial Absorbers: The Effect of Hierarchical Pattern Arrays *Adv. Funct. Mater.* **29**

[51] Aw M S, Simovic S, Yu Y, Addai-Mensah J and Losic D 2012 Porous silica microshells from diatoms as biocarrier for drug delivery applications *Powder Technol.* **223** 52–8

[52] Uthappa U T, Brahmkhatri V, Sriram G, Jung H Y, Yu J, Kurkuri N, Aminabhavi T M, Altalhi T, Neelgund G M and Kurkuri M D 2018 Nature engineered diatom biosilica as drug delivery systems *J. Control. Release* **281** 70–83

[53] Musenich L, Stagni A, Derin L and Libonati F 2024 Tunable Energy Absorption in 3D-Printed Data-Driven Diatom-Inspired Architected Materials *ACS Mater. Lett.*

[54] Perricone V, Santulli C, Rendina F and Langella C 2021 Organismal design and biomimetics: A problem of scale *Biomimetics* **6** 56

[55] Luo S and Greer J R 2018 Bio-Mimicked Silica Architectures Capture Geometry, Microstructure, and Mechanical Properties of Marine Diatoms *Adv. Eng. Mater.* **20** 1–9

[56] Mitchell J G, Seuront L, Doubell M J, Losic D, Voelcker N H, Seymour J and Lal R 2013 The Role of Diatom Nanostructures in Biasing Diffusion to Improve Uptake in a Patchy Nutrient Environment *PLoS One* **8**

[57] Hale M S and Mitchell J G 2002 Effects of Particle Size, Flow Velocity, and Cell Surface Microtopography on the Motion of Submicrometer Particles over Diatoms *Nano Lett.* **2** 657–63

[58] Rosengarten G 2009 Can We Learn From Nature to Design Membranes? The Intricate Pore Structure of the Diatom *ASME 2009 7th International Conference on Nanochannels, Microchannels and Minichannels* (ASMEDC) pp 1371–8

[59] Losic D, Pillar R J, Dilger T, Mitchell J G and Voelcker N H 2007 Atomic force microscopy (AFM) characterisation of the porous silica nanostructure of two centric diatoms *J. Porous Mater.* **14** 61–9

[60] Schmid A M and Volcani B E 1983 WALL MORPHOGENESIS IN COSCINODISCUS WAILESII GRAN AND ANGST. I. VALVE MORPHOLOGY AND DEVELOPMENT OF ITS

|  |  |
|---|---|
| | ARCHITECTURE 1 *J. Phycol.* **19** 387–402 |
| [61] | Zhang J and Ashby M F 1992 The out-of-plane properties of honeycombs *Int. J. Mech. Sci.* **34** 475–89 |
| [62] | Hales T C 2001 The honeycomb conjecture *Discret. Comput. Geom.* **25** 1–22 |
| [63] | Bhatta H, Kong T K and Rosengarten G 2009 Diffusion through diatom nanopores *J. Nano Res.* **7** 69–74 |
| [64] | Buehler M J 2023 Diatom-inspired architected materials using language-based deep learning: Perception, transformation and manufacturing 1–9 |
| [65] | Wu W, Owino J, Al-Ostaz A and Cai L 2014 Applying Periodic Boundary Conditions in Finite Element Analysis *SIMULIA community Conf. Provid.* |
| [66] | Gibson, Lorna J., Ashby M F 1988 *Cellular Solids: Structure and Propertes.* (Cambridge University Press) |
| [67] | Fan X, Verpoest I and Vandepitte D 2006 Finite element analysis of out-of-plane compressive properties of thermoplastic honeycomb *J. Sandw. Struct. Mater.* **8** 437–58 |
| [68] | Piscopo V 2010 Refined buckling analysis of rectangular plates under uniaxial and biaxial compression *World Acad. Sci. Eng. Technol.* **46** 554–61 |

# Supporting Information (SI) file

**Parametric function definition for the flexural moment of inertia**

From Euler-Bernoulli beam theory, it is known that the flexural stiffness $k_{bend}$ of a beam in bending is given by:

$$k_{bend} = \frac{48EI}{L^3}$$

where $E$ is the Young's modulus of the material constituting the beam, $I$ the moment of inertia relative to its bending plane and $L$ its length. Since the dimensions of the beam models and the material used in both simulations and physical tests are always the same, $k_{bend}$ can be considered dependent only on the moment of inertia $I$ and expressed as a function of the geometric parameter affected by the strategy in question as:

$$k_{bend} \propto I(\Delta_{ar})$$

where $\Delta_{ar}$ is the percent change in areolae cell wall thickness.

Given the complicated geometry of the biomimetic structure, for the calculation of $I$ we consider a section of the RVE according to one of its symmetry planes and define a symmetrical IPE beam-type section equivalent to it to represent explicitly through a function the effect of $\Delta_{ar}$ (see Figure S1). Specifically, we associate $\Delta_{ar}$ with the geometric dimensions of the core of the new profile $a_2(\Delta_{a_2})$ and $b_2(\Delta_{b_2})$ and impose that the geometric dimensions of the upper fin ($a_1$ and $b_1$) are equal to those of the lower fin ($a_3$ and $b_3$). In this way the center of gravity G of the new section coincides with its center of symmetry, located at the center of the section. Then, to ensure equivalence, we impose that the flexural moments of inertia of the original section $I_{bio}$ and the dummy section $I_{eq}$ are equal:

$$I_{bio} = I_{eq}$$

We compute $I_{eq}$ as the difference between the moment of inertia $I_{eq_{full}}$ of the rectangle of size $A\text{x}B$ defined by the section's footprint in the plane and the contribution of the voids $I_{eq_{void}}$ that we need to cut from the latter to derive the IPE-type profile (see Figure S1(b)):

$$I_{eq} = I_{eq_{full}} - 2I_{eq_{void}} = \frac{AB^3}{12} - \frac{2}{12} * \left[\frac{A - a_2(\Delta_{a_2})}{2}\right] * [b_2(\Delta_{b_2})]^3$$

The terms $a_2(\Delta_{a_2})$ and $b_2(\Delta_{b_2})$, can be made explicit as a function of the percentage changes in their respective initial values as:

$$a_2(\Delta_{a_2}) = a_2 + \Delta_{a_2} * a_2$$
$$b_2(\Delta_{b_2}) = b_2 + \Delta_{b_2} + b_2$$

Thus, substituting $a_2$ and $b_2$ yields:

$$I_{eq} = \frac{AB^3}{12} - \frac{2}{12} * \left[\frac{A - (a_2 + \Delta_{a_2} a_2)}{2}\right] [b_2 + \Delta_{b_2} b_2]^3$$

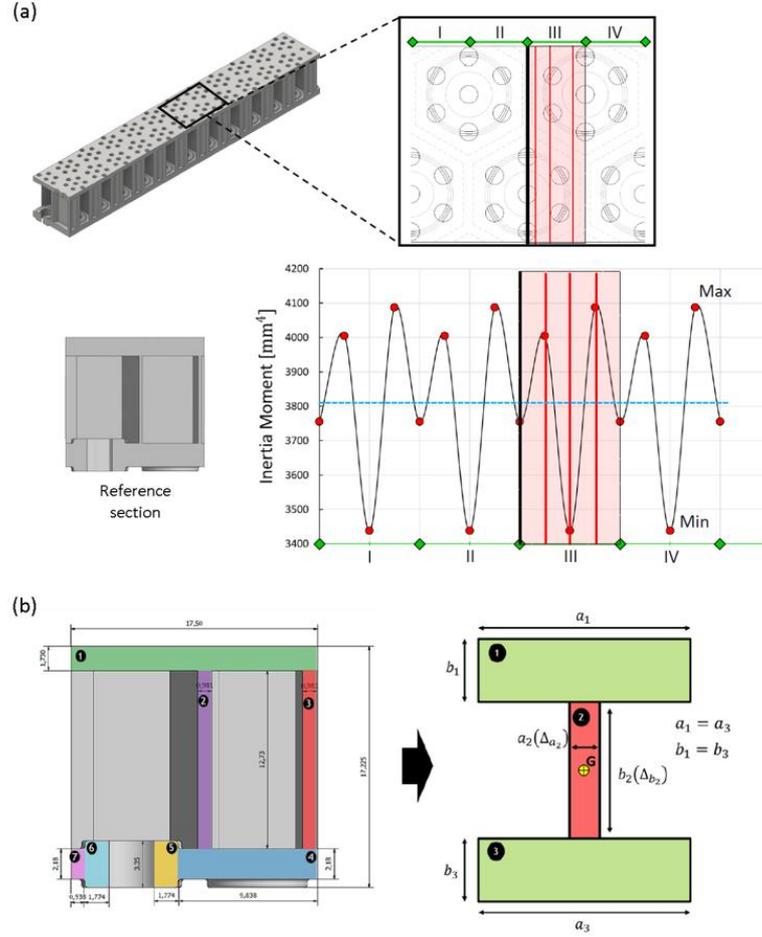

*Figure S1 – (a) Variation of bending moment of inertia as a function of biomimetic model section. The thicker black line represents the plane of symmetry of the RVE taken as a reference to section the structure and compute its geometric properties. In contrast, the red lines represent other section planes chosen to represent the geometric variability associated with the biomimetic model. (b) Transformation of the geometric reference section into an IPE-type profile equivalent to it.*

The source dimension is represented by the side wall thickness of the areolae $a_2$, consequently when a percentage variation $\Delta_{a_2}$ is applied, all other dimensions of the RVE, including $b_2$, vary according to a specific and fixed percentage variation as schematized in Figure S2. Therefore, the variation $\Delta_{b_2}$ can be expressed as:

$$\Delta_{b_2} = C\Delta_{a_2}$$

thus, the expression of $I_{eq}$ can be rewritten as:

$$I_{eq} = \frac{AB^3}{12} - \frac{2}{12} * \left[\frac{A - (a_2 + \Delta_{a_2}a_2)}{2}\right][b_2 + C\Delta_{a_2}b_2]^3$$

representing a one-variable equation as a function of variation $\Delta_{a_2}$, with three unknown constants $A, B, C$.

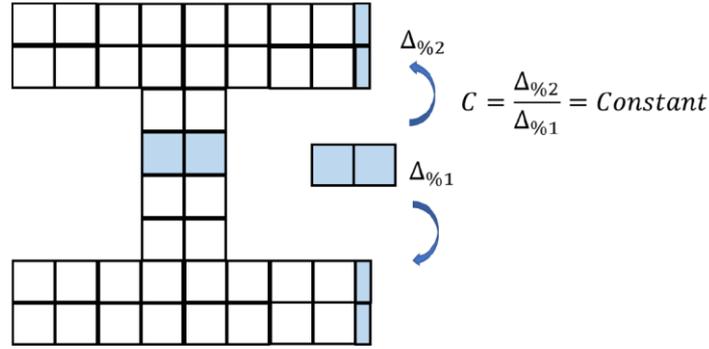

*Figure S2 - Representation of constant percentage variation of dimensions due to the applied constraints to the study.*

The equivalent section has already been chosen to be symmetric and I-shaped, but it is possible to impose a relationship between the two global dimensions A and B to reduce the number of unknowns to be defined. We choose an equivalent geometry with a square cross section, so:

$$A = B$$

By imposing the following system:

$$\begin{cases} A = B \\ I_{bio} = I_{eq}(\Delta a_2 = 0) = \frac{AB^3}{12} - \frac{2}{12}\left[\frac{A-a_2}{2}\right][b_2]^3 \end{cases}$$

we find A = B = 16.863 mm.

The constant C indicates the change in $\Delta b_2$ based on the change in input $\Delta a_2$. Since $C$ does not remain constant for all cases; it is calculated with a set of points corresponding to different $\Delta a_2$, from which the respective $\Delta b_2$ is estimated by the following formula:

$$\Delta_{b_{2_i}} = \frac{b_{2_i}}{b_2} * 100 - 100$$

We use a set of 7 different values of $\Delta a_2$, and the results are shown in Table S1 with the respective values of C we obtained. From these calculations, we define C=0.215.

Figure S3 shows the comparison between the values of the flexural moment of inertia that can be calculated by the classical theoretical approach ($I_{bio}$) and the equivalent formula derived ($I_{eq}$). As it can be seen, the results are practically coincident, so the derived formula successfully approximates the actual moment of inertia of the geometric reference section of the biomimetic model.

*Table S1 - Different versions of the bioinspired samples used to evaluate C: in the first column we report the name of each version used to evaluate different values of C, in the second column the percentage variations of areolae side walls imposed, the third column shows the percentage variation of the height of the areolae side walls (b2) according to the third strategy related to thickness changes, and the last column shows the correspondent values of C obtained in every case. The differences are due to approximations.*

| Version | %$\Delta_{a_2}$ | %$\Delta_{b_2}$ | C |
| --- | --- | --- | --- |
| V_8 | -59.2 | -11.7146 | 0.1979 |
| V_9 | -50 | -10.0327 | 0.2007 |
| V_13 | -20 | -4.216 | 0.2108 |
| V_1 | 0 | 0 | - |
| V_10 | 10 | 2.2238 | 0.2224 |
| V_11 | 20 | 4.5256 | 0.2263 |
| V_12 | 40 | 9.3772 | 0.2344 |

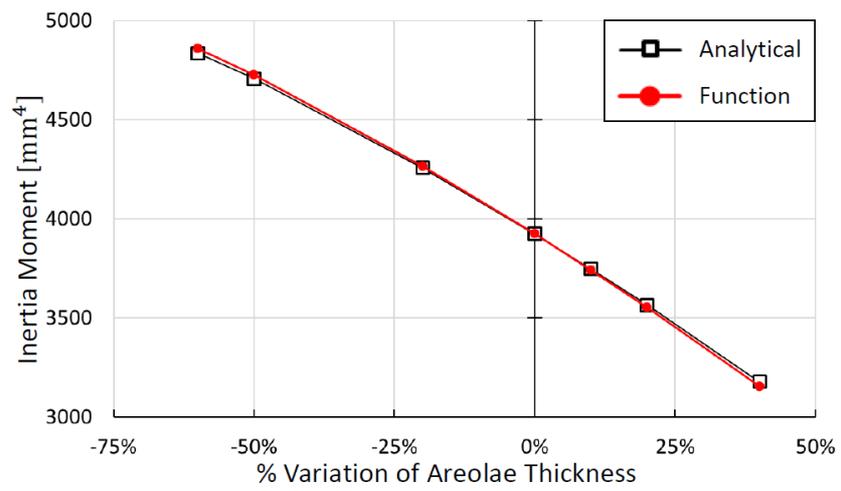

*Figure S3 - Comparison between the values of the flexural moment of inertia that can be calculated by the classical analytical approach and the equivalent formula derived.*

# Details of the biomimetic model in 1:1 scale

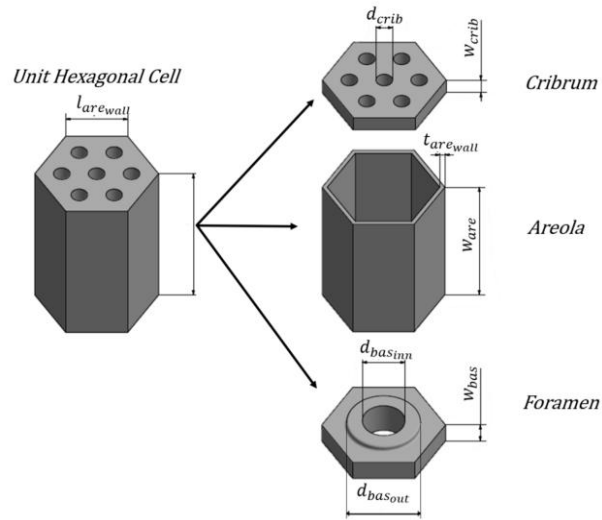

*Figure S4 - Representative Volume Element (RVE) of the diatom-inspired model and respective layer units from which it is composed by.*

*Table S2 - Measurements of geometric dimensions shown in SI Figure 1, obtained from five diatom samples in [32].*

| Spec. | $d_{crib}$ [$\mu m$] | $d_{bas_{inn}}$ [$\mu m$] | $d_{bas_{out}}$ [$\mu m$] | $w_{crib}$ [$\mu m$] | $w_{are}$ [$\mu m$] | $w_{bas}$ [$\mu m$] | $w_{rim}$ [$\mu m$] | $t_{are_{wall}}$ [$\mu m$] | $l_{are_{wall}}$ [$\mu m$] |
|---|---|---|---|---|---|---|---|---|---|
| 1 | 0.34 | 0.8 | 1.47 | 0.42 | 2.66 | 0.62 | 0.29 | 0.17 | 1.19 |
| 2 | 0.34 | 0.82 | 1.37 | 0.26 | 2.97 | 0.41 | 0.19 | 0.17 | 1.19 |
| 3 | 0.35 | 0.89 | 1.49 | 0.36 | 2.67 | 0.48 | 0.28 | 0.17 | 1.19 |
| 4 | 0.27 | 0.9 | 1.43 | 0.31 | 1.94 | 0.32 | 0.2 | 0.17 | 1.19 |
| 5 | 0.3 | 0.7 | 1.33 | 0.38 | 2.49 | 0.35 | 0.21 | 0.17 | 1.19 |

*Table S3 - Averaged values from Table S4 used to design the 3D CAD biomimetic model.*

| Spec. | $d_{crib}$ [$\mu m$] | $d_{bas_{inn}}$ [$\mu m$] | $d_{bas_{out}}$ [$\mu m$] | $w_{crib}$ [$\mu m$] | $w_{are}$ [$\mu m$] | $w_{bas}$ [$\mu m$] | $w_{rim}$ [$\mu m$] | $t_{are_{wall}}$ [$\mu m$] | $l_{are_{wall}}$ [$\mu m$] |
|---|---|---|---|---|---|---|---|---|---|
| Avg | 0.32 | 0.822 | 1.418 | 0.346 | 2.546 | 0.436 | 0.234 | 0.17 | 1.19 |

# Tensile test results

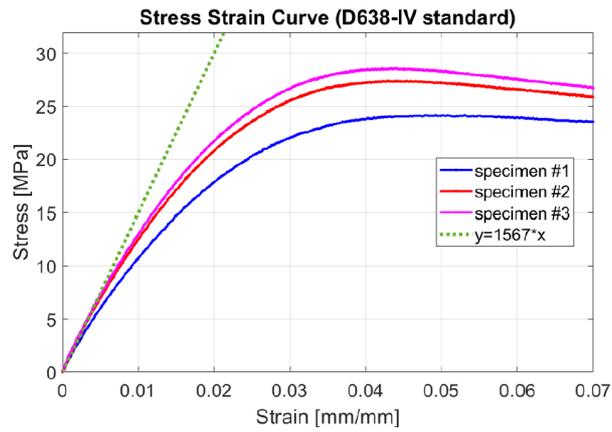

*Figure S5 - Experimental stress-strain curves obtained from three tensile tests on dog-bone samples in VeroBlack®. The slope of the dotted line represents the measured elastic modulus.*

*Table S4 - Main elastic properties obtained from the three dog-bone samples tested.*

| sample ID | Young's Modulus [MPa] | Yield Stress [MPa] | Yield Strain [mm/mm] | Elongation at break [%] |
|---|---|---|---|---|
| D638-IV_1 | 1461 | 24,16 | 0,04785 | 50,7 |
| D638-IV_2 | 1581 | 27,38 | 0,04235 | 44,6 |
| D638-IV_3 | 1658 | 28,55 | 0,04319 | 50,9 |
| Mean Value | 1566,67 | 26,70 | 0,04446 | 48,73 |
| Standard Deviation | 99,27 | 2,27 | 0,00296 | 3,580 |
| Final Value | 1567±99 MPa | 26,7±2,3 MPa | 0,0445±0,003 mm/mm | 48,7±3,6 % |